\documentclass[lettersize,journal]{IEEEtran}
\IEEEpubidadjcol
\usepackage{amsmath,amsfonts}
\usepackage{algorithmic}
\usepackage{algorithm}
\usepackage{array}
\usepackage[caption=false]{subfig}
\usepackage{textcomp}
\usepackage{stfloats}
\usepackage{url}
\usepackage{verbatim}
\usepackage{graphicx}
\usepackage{cite}
\usepackage{multirow}
\usepackage{makecell}
\usepackage{threeparttable}
\usepackage{pifont}
\usepackage{balance}
\usepackage[colorlinks,
	linkcolor=blue,
	anchorcolor=blue,
	citecolor=blue,
	urlcolor=blue]{hyperref}
\usepackage{tikz,xcolor}

\definecolor{lime}{HTML}{A6CE39}
\DeclareRobustCommand{\orcidicon}{
\begin{tikzpicture}
\draw[lime, fill=lime] (0,0)
circle[radius=0.16]
node[white]{{\fontfamily{qag}\selectfont \tiny \.{I}D}};
\end{tikzpicture}
\hspace{-2mm}
}
\foreach \x in {A, ..., Z}{%
\expandafter\xdef\csname orcid\x\endcsname{\noexpand\href{https://orcid.org/\csname orcidauthor\x\endcsname}{\noexpand\orcidicon}}
}

\begin{document}
\title{GA2MIF: Graph and Attention Based Two-Stage Multi-Source Information Fusion for Conversational Emotion Detection}
\author{Jiang Li\hspace{-1.5mm}\orcidA{}, Xiaoping Wang\hspace{-1.5mm}\orcidB{},~\IEEEmembership{Senior Member,~IEEE}, Guoqing Lv, and Zhigang~Zeng\hspace{-1.5mm}\orcidC{},~\IEEEmembership{Fellow,~IEEE}\\
\thanks{Manuscript received 13 September 2022; revised 30 January 2023; accepted 21 March 2023. This work was supported in part by the National Natural Science Foundation of China under Grant 62236005, 61876209, and 61936004. Recommended for acceptance by E. Mower Provost. \textit{(Corresponding authors: Jiang Li and Xiaoping Wang.)}}
\thanks{The authors are with the School of Artificial Intelligence and Automation and the Key Laboratory of Image Processing and Intelligent Control of Education Ministry of China, Huazhong University of Science and Technology, Wuhan 430074, China (e-mail:lijfrank@hust.edu.cn; wangxiaoping@hust.edu.cn; guoqinglv@hust.edu.cn; zgzeng@hust.edu.cn).}
\thanks{Digital Object Identifier 10.1109/TAFFC.2023.3261279}}

\markboth{IEEE TRANSACTIONS ON AFFECTIVE COMPUTING}%
{Li \MakeLowercase{\textit{et al.}}: Graph and Attention Based Two-Stage Multi-Source Information Fusion}

\IEEEpubid{1949--3045~\copyright~2023 IEEE. Personal use is permitted, but republication/redistribution requires IEEE permission.}

\IEEEtitleabstractindextext{%
\begin{abstract}
Multimodal Emotion Recognition in Conversation (ERC) plays an influential role in the field of human-computer interaction and conversational robotics since it can motivate machines to provide empathetic services. Multimodal data modeling is an up-and-coming research area in recent years, which is inspired by human capability to integrate multiple senses. Several graph-based approaches claim to capture interactive information between modalities, but the heterogeneity of multimodal data makes these methods prohibit optimal solutions. In this work, we introduce a multimodal fusion approach named Graph and Attention based Two-stage Multi-source Information Fusion (GA2MIF) for emotion detection in conversation. Our proposed method circumvents the problem of taking heterogeneous graph as input to the model while eliminating complex redundant connections in the construction of graph. GA2MIF focuses on contextual modeling and cross-modal modeling through leveraging Multi-head Directed Graph ATtention networks (MDGATs) and Multi-head Pairwise Cross-modal ATtention networks (MPCATs), respectively. Extensive experiments on two public datasets (i.e., IEMOCAP and MELD) demonstrate that the proposed GA2MIF has the capacity to validly capture intra-modal long-range contextual information and inter-modal complementary information, as well as outperforms the prevalent State-Of-The-Art (SOTA) models by a remarkable margin.
\end{abstract}

}

\maketitle

\IEEEdisplaynontitleabstractindextext

\IEEEpeerreviewmaketitle

\section{Introduction}\label{sec:introduction}
\IEEEPARstart{I}{n} recent years, human-computer interaction and intelligent robotics technologies are transforming science fiction into reality. However, there are numerous challenges to make machines interact with people in a natural manner. The ability to make machines empathize like humans, i.e., to recognize the emotional states of others and respond correspondingly, is particularly crucial in the field of social robotics \cite{leite2013influence}. In addition, empathy can enhance the interaction between human and computer to provide superior artificial intelligence services to others. Accurately recognizing the emotional states of others is a prerequisite for generating empathic responses, which is also a core research thrust in the field of cognition and behavior. Thus, emotion recognition plays an instrumental role in numerous domains and has attracted extensive attention from research scholars.

\IEEEpubidadjcol
Emotion Recognition in Conversation (ERC), also called conversational emotion detection, aims to detect emotional state of a speaker based on the signals that he or she expresses (e.g., the signals include text, audio, or facial expressions). ERC has potential applications in many fields, such as: $(a)$ Disease Diagnosis \cite{bhavan2019bagged}, assisting the doctor in diagnosing disease by identifying the emotional state of the patient when talking to a psychologist. $(b)$ Opinion Mining \cite{cortis2021over}, improving the public's trust in government departments or institutions by analyzing online public opinion and measuring the public's experience of policies or services. $(c)$ Conversation Generation \cite{liang2021infusing}, enhancing significantly the usability of dialogue systems and the satisfaction of customers through injecting emotions into the given model. $(d)$ Recommender System \cite{rosa2018knowledge}, inferring the user's potential preferences by identifying his or her emotional states during historical chats with customer service. ERC systems can provide customized services to users and enhance the quality of empathetic interactions with users.

Most of existing ERC models mainly employ text-modal data as input. DialogueGCN \cite{ghosal2019dialoguegcn} constructs the conversation as a graph to extract long-distance contextual information, where each utterance is related to surrounding utterances. HiGRU \cite{jiao2019higru} adopts lower-level and upper-level Gated Recurrent Units (GRUs) to tackle dilemmas in utterance-level conversational emotion recognition. COSMIC \cite{ghosal2020cosmic} leverages different elements of external commonsense knowledge such as mental states, events, and causal relations to detect utterance-level emotion in conversation. DialogueCRN \cite{hu2021dialoguecrn} attempts to understand conversational context by exploring cognitive factors, which analogous to the unique cognitive thinking of human. Several efforts on modeling based on data from acoustic modalities are also available. Gat et al. \cite{gat2022speaker} introduce a gradient-based adversary learning model that is effective in both speaker-independent and speaker-dependent situations for speech emotion recognition task. Jalal et al. \cite{jalal2020empirical} propose a speech emotion recognition approach based on both Long Short-Term Memory (LSTM) network and Convolutional Neural Network (CNN) to explore the impact of acoustic cues on recognition results. These techniques, anyway, only accept unimodal signal sources as inputs, which may limit the performance of the model. For instance, the model will have trouble properly recognizing emotion if the signal and emotional state of current modality do not match.

\begin{figure}[htbp]
    \centering
    \includegraphics[width=3.4in]{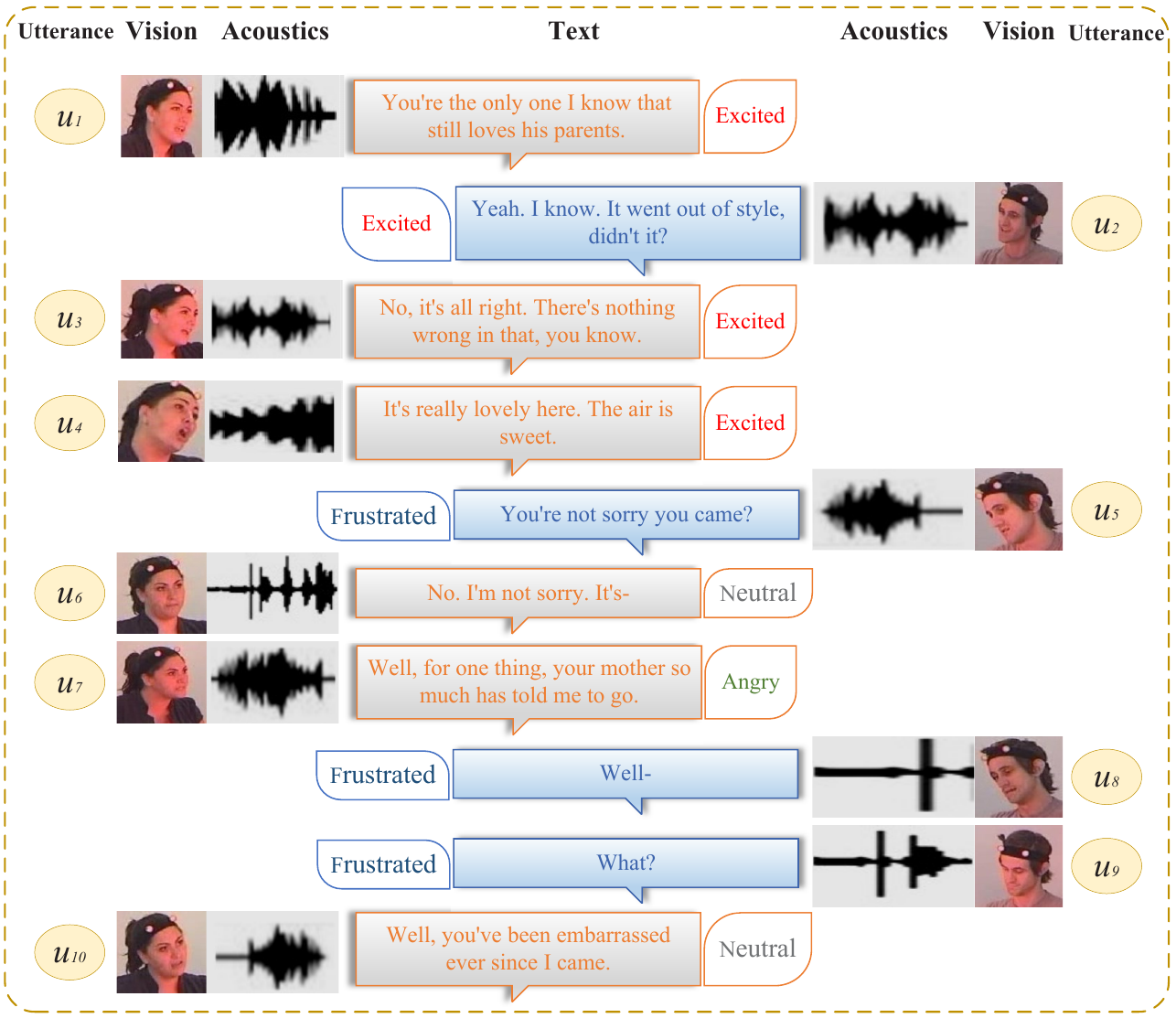}
    \caption{An example of a multimodal conversation scenario, in which the conversational contents are from textual, acoustic and visual modalities. These emotions are the labels of the utterance in the training set; while in the test set, these emotions are the results that are recognized by the model.}
    \label{fig:multi_con}
\end{figure}
Humans are capable of multi-sensory integration, i.e., they can perceive surroundings from multiple senses. Intuitively, multimodal data from different sources can enhance the performance of machine learning models. Fig.~\ref{fig:multi_con} shows a multimodal conversation scene, where the conversational contents are derived from text, acoustic and visual modalities. Multimodal ERC is a burgeoning field of research and has recently been gaining momentum. There are, however, just a few multi-modality based conversational emotion recognition models. BC-LSTM \cite{poria2017context} proposes a multimodal model based on a two-directional LSTM for extracting the contextual information of the current utterance. CMN \cite{hazarika2018conversational} conducts temporal sequence modeling on utterance histories by GRUs and employs attention networks to select the most useful historical utterances. DialogueRNN \cite{majumder2019dialoguernn} is a Recurrent Neural Network (RNN) based ERC framework that considers the characteristic of speaker for each utterance to provide more reliable contextual information. Nevertheless, these methods only concatenate multimodal features in a direct manner, which results in the inter-modal information not being able to interact. Hu et al. \cite{hu2021mmgcn} propose a graph-based multimodal ERC model called MMGCN that claims to effectively not only exploit multimodal dependencies, but also model inter- and intra-speaker dependencies. Experimental results show that MMGCN achieves excellent performance on two public benchmark datasets. Although Graph Neural Networks (GNNs) show excellent performance in homogeneous graphs, they usually obtain suboptimal or even worse results in heterogeneous networks. MMGCN constructs a large graph by treating utterances from different modalities as nodes of the same type, which contradicts the premise that the input to GNN is a homogeneous graph. MMGCN, moreover, simply connects all utterances in each modality to build a complete graph, which undoubtedly brings too much noise (i.e., useless connections) to the model.

In order to address the above-mentioned problems, we propose a Graph and Attention based Two-stage Multi-source Information Fusion (GA2MIF) approach for conversational emotion recognition. The proposed GA2MIF is a new multimodal fusion framework whose network structure mainly consists of graph attention networks and multi-head attention networks. First, we adopt three Multi-head Directed Graph ATtention networks (MDGATs) to extract intra-modal local and long-range contextual information for each modality; then, we leverage three Multi-head Pairwise Cross-modal ATtention networks (MPCATs) to model cross-modal interactions in pairs and extract inter-modal complementary information. Our contributions are mainly as follows:
\begin{enumerate}
\item In this paper, we propose a novel multimodal ERC method named Graph and Attention based Two-stage Multi-source Information Fusion (GA2MIF), which mainly consists of MDGATs and MPCATs. Our GA2MIF not only avoids the use of heterogeneous graph as input to the graph model, but also captures the interaction information between modalities through multi-head attention mechanisms.
\item MDGATs construct three graphs as inputs by treating the utterance of each modality as nodes and the connection of current utterance with a certain range of contextual utterances as edges, thus modeling intra-modal context. This strategy enables the usage of context windows to construct graph instead of fully connected graph, thereby eliminating complex redundant connections. MPCATs model cross-modal interactions by computing the outputs of three modalities of MDGATs with multi-head pairwise attention. Experiments demonstrate that MDGATs can effectively capture intra-modal local and long-range contextual information; MPCATs can effectively capture inter-modal complementary information.
\item We conduct extensive experiments on two public ERC datasets (i.e., IEMOCAP and MELD) and the results show that GA2MIF achieves optimal performance with the accuracy of 69.75\% and 61.65\%.
\end{enumerate}

\section{Related Works}
\subsection{Emotion Recognition}
Emotion recognition is an interdisciplinary realm that has attracted active research and attention in the areas of emotion understanding systems, opinion mining, and emotion generation. Broadly speaking, existing works in this field can be divided into two categories based on data sources: unimodal-based emotion recognition and multimodal-based emotion recognition.

\subsubsection{Unimodal-Based Emotion Recognition}
Unimodal emotion recognition is to judge emotional state of object by encoding input data from single modality, which can be divided into primarily three types: text-based methods, audio-based methods and vision-based methods.

\textit{Text-based Methods} have been the most prevalent unimodal emotion recognition since the development of natural language processing. DialogueGCN \cite{ghosal2019dialoguegcn} models self- and inter-speaker dependencies between speakers to promote context understanding for utterance-level sentiment analysis in conversations. Ishiwatari et al. \cite{ishiwatari2020relation} not only models the dependency of the speaker, but also models the sequential information through relational position encoding. Aiming to model conversational data, DialogXL \cite{shen2021dialogxl} replaces self-attention in XLNet with conversation-aware self-attention to extract the information of intra- and inter-speaker dependencies. DAG-ERC \cite{DBLP:conf/acl/ShenWYQ20} designs a directed acyclic graph neural network to recognize emotion in conversation, where nodes represent utterances and edges represent connections between utterances. HiGRU \cite{jiao2019higru} takes advantage of the lower-level GRU to learn individual utterance embeddings and the upper-level GRU to capture contexts of utterances. COSMIC \cite{ghosal2020cosmic} takes different elements of commonsense into account and makes use of them to learn interactions between speakers. DialogueCRN \cite{hu2021dialoguecrn} designs multi-turn reasoning modules to perceive and combine clues, which can sufficiently extract speaker-level and situation-level context information. \textit{Audio-based Methods}, often referred to as Speech Emotion Recognition (SER), estimate a speaker's emotional state by means of analyzing his/her speech. Gat et al. \cite{gat2022speaker} introduce a general framework for normalizing the features of speakers while addressing the problem of small dataset settings. Jalal et al. \cite{jalal2020empirical} argue that smaller acoustic contexts are crucial in expressing emotion and propose a bidirectional LSTM- and CNN- based SER approach. Guo et al. \cite{guo2021representation} propose a spectro-temporal-channel attention module that offers different weights for frequency, time and channel-wise features to capture more expressive information. \textit{Vision-based Methods} focus on emotion recognition based on facial expression, which is an essential field in affective computing. Jeong et al. \cite{jeong2018driver} propose a fast facial emotion recognition method for recognizing a driver's emotion in real-time. Wang et al. \cite{wang2018intelligent} leverage stationary wavelet entropy to extract features and employ the Jaya algorithm to train facial emotion recognition model. Khaireddin et al. \cite{khaireddin2021facial} fine-tune the hyperparameters of VGGNet architecture to achieve the highest single-network classification accuracy on the FER2013 dataset.

\subsubsection{Multimodal-Based Emotion Recognition}
Unimodal information is insufficient and is easily affected by external factors, such as blocked facial expressions and disturbed voice. In view of complementarity between different modalities, research on multimodal emotion recognition has received increasing attention \cite{poria2017review}. MFN \cite{zadeh2018memory} makes use of a new neural structure based on multi-view sequence learning to consider both view-specific interactions and cross-view interactions. BC-LSTM \cite{poria2017context} proposes a multimodal fusion method that captures contextual information of the utterance through a LSTM network. CMN \cite{hazarika2018conversational} fuses audio, visual and textual features, and it leverages a GRU to model contextual information about historical conversations. ICON \cite{hazarika2018icon} models contextual information through GRU-based memory networks and considers the influence of both self-speaker and inter-speaker. ConGCN \cite{zhang2019modeling} symbolizes the entire conversational corpus as a heterogeneous graph in which each node presents a speaker or an utterance, where each conversation includes textual and acoustic features. DialogueRNN \cite{majumder2019dialoguernn} tracks the states of speakers throughout the conversation by utilizing multiple RNNs for multimodal emotion classification. Relational Tensor Network \cite{sahay2018multimodal} considers relations and interactions of the context segment in a video and shows excellent performance. GME-LSTM(A) \cite{chen2017multimodal} fuses multimodal information at word-level and proposes a model suitable for complex speech structure with gating mechanism to select word-level fusion. MMGCN \cite{hu2021mmgcn} constructs a big graph, which not only captures intra- and inter-speaker dependencies, but also models multimodal information.

\subsection{Machine Learning Methods}
A great number of machine learning applications have made a surge of achievements in recent years relying on Graph Neural Networks (GNNs) and Multi-Head Attention mechanism. In this work, GNNs and multi-head attention mechanism are implemented for intra-modal contextual modeling and inter-modal complementary modeling, respectively.

\subsubsection{Graph Neural Networks}
Graph is a general data representation method describing complex relationship between entities in real scenarios, and has been applied broadly in the industry. However, deep learning has been incapable of effectively adapting to graph structured data. To this end, Graph Neural Networks (GNNs) are proposed to address the above-mentioned challenges. GNNs have been widely applied in numerous fields, including Computer Vision \cite{garcia2017few}, Recommender System \cite{ying2018graph}, Chemistry \cite{gilmer2017neural}, Natural Language Processing \cite{marcheggiani2017encoding}. Given a graph $G=(V,E)$, according to Message Passing Neural Network (MPNN) \cite{gilmer2017neural}, and the information for the $l$-th layer $G$ is updated as follows:
\begin{equation}
    \label{eq:pass_update}
    \begin{split}
    &\mathrm{x}_{v}^{(l+1)} = \mathrm{UPDAT}(\mathrm{x}_{v}^{(l)}, \mathrm{PASS}(\{\mathrm{x}_{u}^{(l)} | u \in \mathcal{N}(v)\})),\\
    &\mathrm{X}_{G} = \mathrm{READOUT}(\{\mathrm{x}_{v}^{(L)}\}, v \in V),       
    \end{split}
\end{equation}
where $l=0,1,\cdots,L$, and $L$ is the number of layers; $\mathrm{x}_{v}^{(l)}$ is the $l$-th layer representation of node $v$, and $\mathrm{x}_{v}^{(0)}$ is initial input of node $v$; $\mathcal{N}(v)$ denotes the neighboring set of node $v$, and $u$ is a neighborhood of node $v$; $\mathrm{PASS}$ and $\mathrm{UPDAT}$ are the parameterized message passing function and state updating function, and $\mathrm{READOUT}$ is the readout function.

The widely used GNN models include 1stChebNet \cite{kipf2016semi}, GraphSAGE \cite{hamilton2017inductive} and GAT \cite{velickovic2017graph}. GAT supposes that contributions of neighboring nodes to current node are neither identical as GraphSage, nor predefined as 1stChebNet. GAT introduces attention mechanism to calculate the importance of neighboring nodes, which is defined as:
\begin{equation}
    \label{eq:gat0}
    \mathrm{x}_{v}^{(l+1)} = \delta \left({\sum _{u\in \mathcal {N}(v)}\mu _{vu}^{(l+1)}\mathrm{W}^{(l+1)}\mathrm {x}_{u}^{(l)}}\right),
\end{equation}
where $\mathrm{x}_{v}^{(l+1)}$ is the $(l+1)$-th layer representation of current node $v$; $u$ are neighboring node of $v$; $\delta$ is nonlinear activation function, and $\mathrm{W}^{(l+1)}$ denotes the trainable parameter. The attention weight $\mu _{vu}^{(l+1)}$ indicates the importance of $u$ to $v$:
\begin{equation}
    \label{eq:attention_weight}
    \mu _{vu}^{(l+1)} = \frac{\exp (\sigma(\mathrm{a}^{T} [\mathrm{W}^{(l+1)}\mathrm{x}_{v}^{(l)} | | \mathrm{W}^{(l+1)}\mathrm{x}_{u}^{(l)}]))}
    {\sum_{w \in \mathcal{N}(v)} \exp (\sigma(\mathrm{a}^{T} [\mathrm{W}^{(l+1)}\mathrm{x}_{v}^{(l)} | | \mathrm{W}^{(l+1)}\mathrm{x}_{w}^{(l)}]))},
\end{equation}
where $\sigma$ is the LeakyReLU function, and $u$, $w$ are neighbors of $v$; both $\mathrm{a}$ and $\mathrm{W}^{(l+1)}$ are the learnable parameters. In addition, multi-head GAT is executed to increase the expressiveness of the model \cite{velickovic2017graph}.

\subsubsection{Multi-Head Attention Mechanism}
Multi-Head Attention mechanism is first proposed in Transformer \cite{vaswani2017attention} architecture, which is inspired by attention model \cite{bahdanau2014neural}. Transformer is a new type of neural network that has been widely adopted in various fields, such as Natural Language Processing \cite{vaswani2017attention,devlin2018bert}, Computer Vision \cite{dosovitskiy2020image,carion2020end}, and Speech Processing \cite{chen2021developing,dong2018speech}. Devlin et al. \cite{devlin2018bert} propose a language representation model named BERT, which received enthusiastic attention once it was proposed due to its excellent performance. ViT \cite{dosovitskiy2020image} implements direct application of a standard Transformer to image classification tasks, and is a classic application that adapts Transformer to the field of computer vision. Dong et al. \cite{dong2018speech} extend Transformer to Automatic Speech Recognition (ASR), and proposed model is named Speech-Transformer. Speech-Transformer can achieve a reduced training cost compared to the majority of recurrence-based models.

Given packed feature representation query $\mathrm{Q}$, key $\mathrm{K}$, value $\mathrm{V}$, the scaled dot-product attention is computed as:
\begin{equation}
    \label{eq:attention}
    \mathrm{Att}(\mathrm{Q},\mathrm{K},\mathrm{V})=\mathrm{softmax}(\frac{\mathrm{Q}\cdot \mathrm{K}^\top}{\sqrt{d_k}})\cdot \mathrm{V},
\end{equation}
where $d_k$ denotes the dimensions of $\mathrm{K}$ or $\mathrm{V}$; $\mathrm{softmax}(\frac{\mathrm{Q}\cdot \mathrm{K}^\top}{\sqrt{d_k}})$ is called attention matrix; $\mathrm{softmax}$ denotes the softmax function, which is employed in a row-wise manner. Multi-Head Attention can be a mechanism that can enhance the stability and performance of the vanilla single attention. Specifically, different heads employ different query, key and value matrices. Multi-head attention can be formalized as follows:
\begin{equation}
    \label{eq:multi_attention}
    \begin{split}
    &\mathrm{MA}(\mathrm{Q},\mathrm{K},\mathrm{V})=\mathrm{W}_{ma}[\mathrm{head}_0 \ \Vert \cdots \Vert \ \mathrm{head}_h],\\
    &\mathbf{s.t.}\ \mathrm{head}_i=\mathrm{Att}(\mathrm{W}_{\mathrm{Q},i} \mathrm{Q}, \mathrm{W}_{\mathrm{K},i} \mathrm{K},\mathrm{W}_{\mathrm{V},i} \mathrm{V}),
    \end{split}
\end{equation}
where $\mathrm{MA}$, $\mathrm{Att}$ are the multi-head attention and single attention function, and $\Vert$ denotes the concatenation operation; $\mathrm{W}_{\mathrm{Q},i}$, $\mathrm{W}_{\mathrm{K},i}$, $\mathrm{W}_{\mathrm{V},i}$ are the learnable parameters, which can project $\mathrm{Q}$, $\mathrm{K}$, $\mathrm{V}$ into different representation subspaces, respectively; $\mathrm{W}_{ma}$ is also the trainable parameter.

\section{Methodology}\label{methodology}
\begin{figure*}[thbp]
    \centering
    \includegraphics[width=7in]{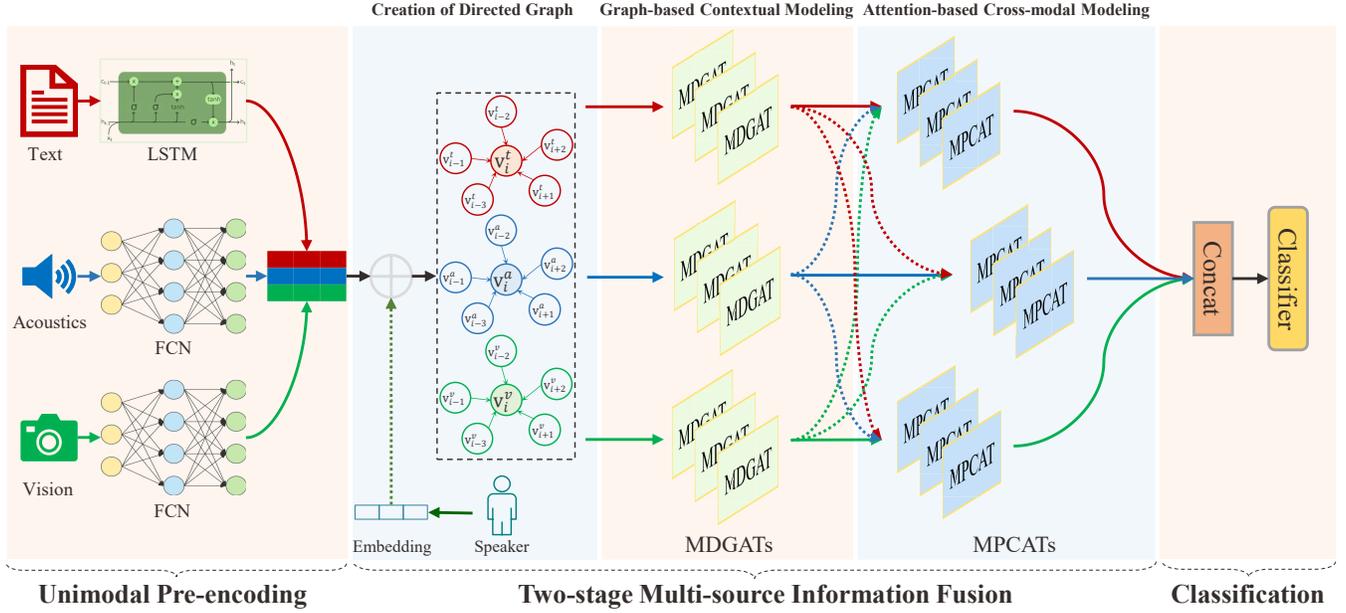}
    \caption{The overall framework of the proposed approach. Here, two-stage multi-source information fusion mainly involves the creation of directed graph, graph-based contextual modeling, and attention-based cross-modal modeling.}
    \label{fig:overall}
\end{figure*}
In this section, we introduce overall framework of the proposed method in detail. Our proposed framework is exhibited in Fig.~\ref{fig:overall}, which consists of unimodal pre-encoding, two-stage multi-source information fusion, and classification.

\subsection{Problem Definitions}
Before presenting overall framework of the proposed approach, several definitions covering the objective task, intra-modal contextual information, and inter-modal complementary information will be given. 

\textit{Objective Task:} A conversation consists of $m$ utterances $u_1$, $u_2$, $\cdots$, $u_m$, and each utterance $u_i$ corresponds to an emotion label $y_i$. Different emotion labels are available for different datasets, e.g., \textit{Happy}, \textit{Sad}, \textit{Neutral}, \textit{Angry}, \textit{Excited} and \textit{Frustrated} for the IEMOCAP dataset. There are more than two participants in a conversation, so $u_i$ and $u_j$ may be spoken by the same speaker or by different speakers. Each utterance $u_i$ has three modal expressions $u_i^t$, $u_i^a$, $u_i^v$ in multimodal ERC, which corresponds to textual, acoustic and visual modalities, respectively. Given an utterance $u_i$, the task of ERC is to take $u_i$ as input and detect the corresponding emotional state $y_i$.

\textit{Intra-modal Contextual Information:} In a modality, the $j$-th utterance $u_j$ before the utterance $u_i$ ($j<i$) is its context. Meanwhile, the context of $u_i$ also includes the $k$-th utterance $u_k$ ($k>i$) after it. Here, the information carried by $u_j$ and $u_k$ are contextual information of $u_i$. Especially, if $j=i-\epsilon$ or $k=i+\epsilon$ and $\epsilon$ is a small integer, then $u_j$ or $u_k$ is local contextual utterance of $u_i$. Conversely, if $\epsilon$ is a large integer, then $u_j$ or $u_k$ is long-range contextual utterance. Commonly, recurrence-based methods have difficulty catching long-range contextual information.

\textit{Inter-modal Complementary Information:} In multiple modalities, the utterance $u_i$ can be represented as $u_i^t$, $u_i^a$, and $u_i^v$. The information carried by $u_i^a$ can be regarded as the complementary information of $u_i^t$. Similarly, $u_i^v$ can be also regarded as the complementary information of $u_i^t$, and the rest may be deduced by analogy.

\subsection{Unimodal Pre-Encoding}
To rigorously demonstrate the superiority of our model over MMGCN, we employ the identical unimodal encoding method as Hu et al. \cite{hu2021mmgcn}. Specifically, we adopt a Bi-directional LSTM (BiLSTM) network to extract the contextual information of textual modality; unlike textual modality, we adopt the fully connected networks to encode acoustic and visual modalities. The unimodal pre-encoding processes for three modalities can be formalized as:
\begin{equation}
    \label{eq:pre_encoding}
    \begin{split}
    &\mathrm{o}_i^\delta = \mathrm{W}_o^\delta \mathrm{u}_i^\delta + \mathrm{b}_{o}^\delta, \delta \in \{a, v\},\\ 
    &\mathrm{o}_i^t, \mathrm{o}_{h,i}^t = \mathrm{BiLSTM}(\mathrm{u}_i^t, \mathrm{o}_{h,i-1}^t, \mathrm{o}_{h,i+1}^t),
    \end{split}
\end{equation}
where $\mathrm{u}_i^t$, $\mathrm{u}_i^a$, and $\mathrm{u}_i^v$ are the inputs of unimodal pre-encoding; $t$, $a$, and $v$ denote textual, acoustic and visual modalities; $\mathrm{o}_i^t$, $\mathrm{o}_i^a$, and $\mathrm{o}_i^v$ are the outputs of unimodal pre-encoding; $\mathrm{o}_{h,i}^t$ contains the $i$-th cell state and hidden state; $\mathrm{W}_o^\delta$ and $\mathrm{b}_{o}^\delta$ are the trainable parameters.

\subsection{Two-Stage Multi-Source Information Fusion}
To adequately capture intra-modal contextual information and inter-modal complementary information, we present a Graph and Attention based Two-stage Multi-source Information Fusion (GA2MIF) technique. First, we model the intra-modal contexts using three Multi-head Directed Graph ATtention networks (MDGATs) that fully capture the contextual information of textual, acoustic and visual modalities. Then, three Multi-head Pairwise Cross-modal ATtention network (MPCATs) are utilized for cross-modal modeling, which allow for inter-modal feature interactions and thus capture complementary information about inter-modality.

\subsubsection{Characteristics of the Speaker}
Different speakers exhibit distinct self-characteristics in a conversation, such as various personalities, timbres, and expressions. Therefore, we believe that the speaker's information is of importance for emotion recognition. In order to extract self-characteristics of speakers, we first encode the speaker to obtain embedding vector, i.e., speaker embedding; and then we add speaker embedding to the corresponding utterance. This process can be expressed as follows:
\begin{equation}
    \label{eq:emb-spk}
    \begin{split}
    &\mathrm{S}_{e} = \mathrm{EMB}(S, n),\\
    &X^\tau = \mathrm{O}^\tau + \lambda \mathrm{S}_{e},
    \end{split}
\end{equation}
where $\mathrm{EMB}$ represents Embedding function; $S$ is the set of speakers, $n$ is the number of speakers, and $\mathrm{S}_{e}$ denotes speaker embedding; $X^\tau$ denotes feature matrix adding speaker embeddings, $\tau \in \{t,a,v\}$, and $\mathrm{O}^\tau$ is feature matrix from uni-modal pre-encoding phase, $\mathrm{o}_i^\tau \in \mathrm{O}^\tau$; $\lambda$ is trade-off parameter of speaker embedding.

\subsubsection{Creation of Directed Graph}
As shown in Fig.~\ref{fig:overall}, we create three directed graphs $G^t$, $G^a$ and $G^v$ for a conversation. $G^t$, $G^a$ and $G^v$ can be represented as $G^\tau=(V^\tau,E^\tau,\mathcal{W}^\tau)$ ($\tau \in \{t,a,v\}$), where $V^\tau$ denotes the set of nodes, $E^\tau$ denotes the set of edges, i.e., the set of connections between nodes, and $\mathcal{W}^\tau$ denotes the set of edge weights. Specifically, in multimodal ERC, our graph is created as follows.

\textit{Nodes:} In a conversation, each utterance $u_i$ is represented as three nodes $\mathrm{v}_i^t$, $\mathrm{v}_i^a$, and $\mathrm{v}_i^v$. Here, $t$, $a$, $v$ denote textual, acoustic, and visual modalities; $\mathrm{v}_i^t \in V^t$, $\mathrm{v}_i^a \in V^a$, $\mathrm{v}_i^v \in V^v$. Given $m$ utterances, we create $3 \times m$ nodes, i.e., $|V^\tau|=3 \times m$, and $m$ is the number of utterances in current conversation.

\textit{Edges:} Assuming that only unimodality is considered, we connect utterance node $\mathrm{v}_i$ in the conversation with its past $\mathcal{J}$ (future $\mathcal{K}$) contextual utterance nodes $\mathrm{v}_{i-\mathcal{J}}$, $\mathrm{v}_{i-\mathcal{J}+1}$, $\cdots$, $\mathrm{v}_{i-1}$ ($\mathrm{v}_{i+1}$, $\mathrm{v}_{i+2}$, $\cdots$, $\mathrm{v}_{i+\mathcal{K}}$). Here, $\mathcal{J}$, $\mathcal{K}$ are defined as the window size of past and future contexts. Based on this, we construct edges using the above-mentioned strategy in the textual, acoustic and visual modalities respectively.

\textit{Edge Weights:} The edge weight can distinguish the importance of different neighboring nodes and is an important element in graph neural networks. Veli{\v{c}}kovi{\'c} et al. \cite{velickovic2017graph} propose Graph ATtention network (GAT) that claim to learn the edge weights of the graph. The attention scores are calculated as follows:
\begin{equation}
    \label{eq:gat}
    \mu_{ij}^\tau =
    \frac{
    \exp\left(\sigma\left({(\mathrm{a}^\tau)}^{\top}
    [\mathrm{W}_\mathrm{ew}^\tau x_i^\tau \, \Vert \, \mathrm{W}_\mathrm{ew}^\tau x_j^\tau]
    \right)\right)}
    {\sum_{\mathrm{v}_k^\tau \in \mathcal{N}(\mathrm{v}_i^\tau)}
    \exp\left(\sigma\left({(\mathrm{a}^\tau)}^{\top}
    [\mathrm{W}_\mathrm{ew}^\tau x_i^\tau \, \Vert \, \mathrm{W}_\mathrm{ew}^\tau x_k^\tau]
    \right)\right)},
\end{equation}
where $x_i^\tau \in X^\tau$ denotes feature vector of node $\mathrm{v}_i^\tau$ in the graph, and $X^\tau$ is feature matrix of $V^\tau$, $\tau \in \{t,a,v\}$; both $x_j^\tau$ and $\mathrm{v}_k^\tau$ are neighbor node of $\mathrm{v}_i^\tau$, and $\mathrm{v}_j^\tau$, $\mathrm{v}_k^\tau$ $\in$ $\{\mathrm{v}_{i-\mathcal{J}}^\tau$, $\cdots$, $\mathrm{v}_{i-1}^\tau$, $\mathrm{v}_{i+1}^\tau$, $\cdots$, $\mathrm{v}_{i+\mathcal{K}}^\tau\}$; $\mu_{ij}^\tau$ is weight of the edge between node $\mathrm{v}_i^\tau$ and $\mathrm{v}_j^\tau$, and also attention coefficient of GNN; $\Vert$ denotes concatenation operation; $\mathrm{\sigma}$ represents non-linear activation function, e.g, $\mathrm{LeakyReLU}$; $\mathrm{a}^\tau$ and $\mathrm{W}_\mathrm{ew}^\tau$ are the learnable parameters. We set the window size of context to get neighbors instead of taking all nodes in the conversation as neighbors, which not only improves computational efficiency but also increases the stability of the model (since the window size of context for each node $\mathrm{v}_i^\tau$ is fixed in each experiment).

In the standard GAT scoring function, however, $\mathrm{W}_\mathrm{ew}^\tau$ and $\mathrm{a}^\tau$ are applied consecutively. This way can be converted into a linear layer, limiting the expressiveness of attention function. In order to address the above dilemma, we employ an attention module based on GATv2 to set edge weights referring to the idea of Brody et al. \cite{brody2022how}. Specifically, our edge weights are calculated as follows:
\begin{equation}
    \label{eq:edge-weights}
    \mu_{ij}^\tau =
    \frac{
    \exp\left((\mathrm{a}^\tau)^{\top}\sigma\left(\mathrm{W}_\mathrm{ew}^\tau
    [x_i^\tau \, \Vert \, x_j^\tau]
    \right)\right)}
    {\sum_{\mathrm{v}_k^\tau \in \mathcal{N}(\mathrm{v}_i^\tau)}
    \exp\left((\mathrm{a}^\tau)^{\top}\sigma\left(\mathrm{W}_\mathrm{ew}^\tau
    [x_i^\tau \, \Vert \, x_k^\tau]
    \right)\right)},
\end{equation}

\begin{figure}[htbp]
    \centering
    \subfloat[MDGAT]{\includegraphics[width=0.9in]{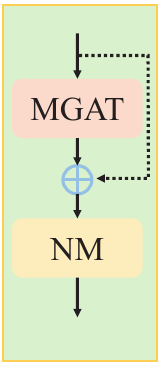}%
    \label{fig:mdgat}}
    \hfil
    \subfloat[MPCAT]{\includegraphics[width=2.5in]{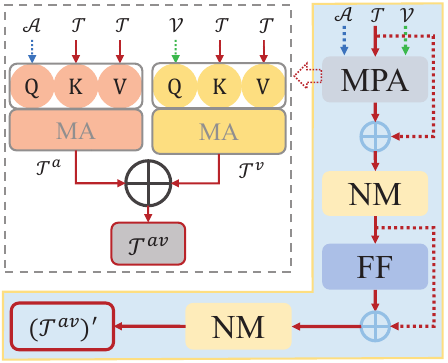}%
    \label{fig:mpcat}}
    \caption{The illustration of MDGAT and MPCAT, where MGAT and MPA denote multi-head graph attention and multi-head pairwise attention (consisting of two multi-head attentions), respectively.}
    \label{fig:mdgat-mpcat}
\end{figure}

\subsubsection{Graph-Based Contextual Modeling}\label{mdgats}
Graph Neural Networks (GNNs) have been proved to have excellent relational modeling capabilities. We employ three Multi-head Directed Graph ATtention networks (MDGATs) to model the contextual dependencies of utterances in the conversation. To alleviate the over-smoothing problem \cite{li2018deeper} of GNNs, we connect the previous layer input to the next layer input. Specifically, the input of the prior layer GNN and its output are first added together; then taking the obtained result as the input of the next layer GNN (see Fig.~\ref{fig:mdgat}). Our designed MDGATs can be formalized as:
\begin{equation}
    \label{eq:mdgat}
    \mathrm{X}^{\tau,l+1} = \mathrm{NM}(\mathrm{X}^{\tau,l} + \mathrm{MGAT}(\mathrm{X}^{\tau,l},E^\tau)),
\end{equation}
where $l=0,1,\cdots,L$, and $L$ is the number of layers of MDGATs; $\mathrm{X}^{\tau,l}$ is the $l$-th layer feature matrix, $\mathrm{X}^{\tau,0}=X^\tau$, and $E^\tau$ is edge set; $\mathrm{NM}$ and $\mathrm{MGAT}$ are the normalization layer and multi-head graph attention layer, respectively. Here, $\mathrm{MGAT}$ can be computed as:
\begin{equation}
    \label{eq:mgat}
    \begin{split}
    &\mathrm{MGAT}(\mathrm{X}^{\tau,l},E^\tau) = \mathrm{W}_{mg}^l[\mathrm{head}_0^l \ \Vert \cdots \Vert \ \mathrm{head}_h^l],\\
    &\mathbf{s.t.}\ \mathrm{head}_i^l=\mathrm{GAT}(\mathrm{X}^{\tau,l},E^\tau),
    \end{split}
\end{equation}
where $\Vert$ is concatenation operation, and $\mathrm{W}_{mg}^l$ is the trainable parameter. We describe the process of a single GAT according to MPNN \cite{gilmer2017neural}, i.e., GAT is divided into two phases: message passing and state updating.

\textit{Message Passing:} The purpose of message passing is to aggregate the information of neighboring nodes. With the help of the calculation of edge weights in Equation \ref{eq:edge-weights}, we distinguish the importance of different nodes when aggregating contextual information. For multimodal ERC, our passing function is defined as follows:
\begin{equation}
    \label{eq:passing}
    \mathrm{x}_{ps,i}^{\tau,l+1} = \sum_{{\mathrm{v}_j^\tau} \in \mathcal{N}(\mathrm{v}_i^\tau)} \mu_{ij}^{\tau,l+1} \mathrm{W}_{ps}^{\tau,l+1} \mathrm{x}_j^{\tau,l},
\end{equation}
where $\mu_{ij}^{\tau,l+1}$ is the $(l+1)$-th layer attention coefficient, as well as the $(l+1)$-th layer edge weight between node $\mathrm{v}_i^\tau$ and $\mathrm{v}_j^\tau$; $\mathrm{v}_j^\tau$ is neighboring node of $\mathrm{v}_i^\tau$, and $i \neq j$; $\mathrm{x}_j^{\tau,l} \in \mathrm{X}^{\tau,l}$ is the $l$-th layer feature vector of $\mathrm{v}_j^\tau$, and $\mathrm{x}_{ps,i}^{\tau,l+1}$ denotes the $(l+1)$-th layer output of message passing; $\mathrm{W}_{ps}^{\tau,l+1}$ denotes the learnable parameter. $\mathrm{x}_j^{\tau,0}$ is initial input feature vector of $\mathrm{v}_j^{\tau}$, i.e., $\mathrm{x}_j^{\tau,0}=x_j^{\tau}$, $\mathrm{X}^{\tau,0}=X^{\tau}$, $\mathrm{x}_j^{\tau,0}\in \mathrm{X}^{\tau,0}$, $x_j^{\tau}\in X^{\tau}$.

\textit{State Updating:} In state updating phase, the updating function combines the $(l+1)$-th layer output ($\mathrm{x}_{ps,i}^{\tau,l+1}$) of message passing with the $l$-th layer vector representation ($\mathrm{x}_i^{\tau,l}$) of node $\mathrm{v}_i^{\tau}$ to obtain the $(l+1)$-th layer vector representation ($\mathrm{x}_i^{\tau,l+1}$) of $\mathrm{v}_i^{\tau}$. Here, $\tau \in \{t,a,v\}$, $l=0,1,\cdots,L$, and $L$ is the number of layers of MDGATs. We adopt a variety of updating functions, and $\mathrm{x}_i^{\tau,l+1}$ can be computed as follows.
\begin{itemize}
	\item \textit{Sum Updating} first applies the linear transformation to each of two vector representations, and then sums obtained results:
	\begin{equation}
        \label{eq:sum_updating}
        \mathrm{x}_{sum,i}^{\tau,l+1} = \mathrm{W}_{sum0}^{\tau,l+1}\mathrm{x}_{ps,i}^{\tau,l+1}+\mathrm{W}_{sum1}^{\tau,l+1}\mathrm{x}_i^{\tau,l},
    \end{equation}
    where $\mathrm{x}_{sum,i}^{\tau,l+1}$ is the $l$-th layer output of state updating, as well as the $(l+1)$-th layer vector representation of $\mathrm{v}_i$; $\mathrm{W}_{sum0}^{\tau,l+1}$ and $\mathrm{W}_{sum1}^{\tau,l+1}$ are the trainable parameters.

	\item \textit{Concat Updating} applies the concatenation operation to the two vector representations, followed by the linear transformation:
	\begin{equation}
        \label{eq:concat_updating}
        \mathrm{x}_{cat,i}^{\tau,l+1} = \mathrm{W}_{cat}^{\tau,l+1}[\mathrm{x}_{ps,i}^{\tau,l+1} \ \Vert \ \mathrm{x}_i^{\tau,l}],
    \end{equation}
    where $\Vert$ is the concatenation operation, and $\mathrm{W}_{cat}^{\tau,l+1}$ is the learnable parameter.

	\item \textit{Sum-Product Updating} is our elaborate updating function. We compute in Sum-Product not only the sum of two vectors, but also their element-wise product, thus considering two types of feature interactions between $\mathrm{x}_{ps,i}^{\tau,l}$ and $\mathrm{x}_i^{\tau,l}$:
	\begin{equation}
        \label{eq:sum_product_updating}
        \begin{split}
        &\mathrm{x}_{sump,i}^{\tau,l+1} =\\
        &\mathrm{W}_{sump}^{\tau,l+1}[(\mathrm{x}_{ps,i}^{\tau,l+1}+\mathrm{x}_i^{\tau,l}) \ \Vert \ (\mathrm{x}_{ps,i}^{\tau,l+1} \odot  \mathrm{x}_i^{\tau,l})],           
        \end{split}
    \end{equation}
    where $\odot$ denotes element-wise product, and $\mathrm{W}_{sump}^{\tau,l+1}$ is the learnable parameter. We hope that \textit{Sum-Product Updating} function can update more messages from similar contextual neighbors to current node $\mathrm{v}_i^\tau$.
\end{itemize}

\subsubsection{Attention-Based Cross-Modal Modeling}
In this part, we introduce three Multi-head Pairwise Cross-modal ATtention networks (MPCATs) for cross-modal feature interaction and thus capture intra-modal complementary information. The structure of MPCAT can be seen in Fig.~\ref{fig:mpcat}. We can observe that the proposed MPCAT mainly consist of two multi-head attention layers ($\mathrm{MA}$ denotes the multi-head attention layer) and one feedforward layer ($\mathrm{FF}$ denotes the feedforward layer), where the input of each MPCAT contains the feature matrices of two modalities. Note that the inputs of MPCATs are based on the outputs of MDGATs. In the following, we describe the process of MPCATs in detail.

Firstly, we treat acoustic feature matrix $\mathcal{A}$ as query $\mathrm{Q}$ of multi-head attention layer, and textual feature matrix $\mathcal{T}$ as key $\mathrm{K}$ and value $\mathrm{V}$ of multi-head attention layer; after the attention calculation, we obtain textual-acoustic interaction matrix $\mathcal{T}^{a}$. Similarly, in order to obtain textual-visual interaction matrix $\mathcal{T}^{v}$, we apply visual feature matrix $\mathcal{V}$ and textual feature matrix $\mathcal{T}$ to another multi-head attention layer; where query $\mathrm{Q}$ is replaced by visual feature matrix $\mathcal{V}$, and key $\mathrm{K}$ and value $\mathrm{V}$ are replaced by textual feature matrix $\mathcal{T}$. Secondly, we add $\mathcal{T}^{a}$ and $\mathcal{T}^{v}$ together; and the obtained result is input to the residual connection layer and normalization layer in turn to get textual feature matrix $\mathcal{T}^{av}$ that are interacted by acoustic and visual modalities; where the residual connection and normalization layer are to ensure the stability of $\mathcal{T}^{av}$. Finally, with reference to network structure of Transformer \cite{vaswani2017attention}, $\mathcal{T}^{av}$ is sequentially applied to the feedforward layer, residual connection layer and normalization layer; where the residual connection and normalization layer to enhance the expressiveness of $\mathcal{T}^{av}$. After all above-mentioned steps, a new textual feature matrix $(\mathcal{T}^{av})'$ encoded by MPCAT is acquired, which carries the information of acoustic and visual modality. We repeat above operation $K$ times ($K$ is the number of layers of MPCATs), then we get final output of MPCATs.

The $k$-th layer feature matrix $\mathcal{T}^{av,k}$ ($k=0,1,\cdots,K$) can be formulated as:
\begin{equation}
    \label{eq:mpcat0_text}
    \begin{split}
    &\mathcal{T}^{av,k}=\mathrm{NM}(\mathcal{T}^k+\mathrm{MPA}(\mathcal{A}^k,\mathcal{V}^k,\mathcal{T}^k)),\\
    &\mathbf{s.t.}\ \mathrm{MPA}(\mathcal{A}^k,\mathcal{V}^k,\mathcal{T}^k)=\\ 
    &\quad\ \ \mathrm{DP}(\mathrm{MA}(\mathcal{A}^k,\mathcal{T}^k,\mathcal{T}^k)+\mathrm{MA}(\mathcal{V}^k,\mathcal{T}^k,\mathcal{T}^k)),
    \end{split}
\end{equation}
where $\mathcal{T}^k$, $\mathcal{A}^k$, and $\mathcal{V}^k$ are the $k$-th layer feature matrices of textual, acoustic, and visual modalities, respectively; $\mathrm{MA}(\mathcal{A}^k,\mathcal{T}^k,\mathcal{T}^k)=\mathcal{T}^{a,k}$, $\mathrm{MA}(\mathcal{V}^k,\mathcal{T}^k,\mathcal{T}^k)=\mathcal{T}^{v,k}$; $\mathrm{NM}$, $\mathrm{DP}$, and $\mathrm{MA}$ denote the normalization, dropout and multi-head attention layers, respectively; $\mathrm{MPA}$ includes a dropout layer and two multi-head attention layers (i.e., two $\mathrm{MA}$). $\mathrm{MA}$ can be represented as:
\begin{equation}
    \label{eq:multiatt}
    \begin{split}
    &\mathrm{MA}(\mathcal{A}^k,\mathcal{T}^k,\mathcal{T}^k)=\mathrm{W}_{ma}^k[\mathrm{head}_0^k \ \Vert \cdots \Vert \ \mathrm{head}_h^k],\\
    &\mathbf{s.t.}\ \mathrm{head}_i^k=\mathrm{Att}(\mathrm{W}_{\mathrm{Q},i}^k \mathcal{A}^k, \mathrm{W}_{\mathrm{K},i}^k \mathcal{T}^k,\mathrm{W}_{\mathrm{V},i}^k \mathcal{T}^k),
    \end{split}
\end{equation}
where $\mathrm{head}_i^k$ is the output of the $i$-th single attention layer, $\mathrm{Att}$ denotes the single attention layer; $\mathrm{W}_{ma}^k$, $\mathrm{W}_{\mathrm{Q},i}^k$, $\mathrm{W}_{\mathrm{K},i}^k$, and $\mathrm{W}_{\mathrm{V},i}$ are the trainable parameters.

The above-mentioned feature matrix $(\mathcal{T}^{av,k})'$ is the output of the $k$-th layer MPCAT, as well as the input of the $(k+1)$-th layer MPCAT, i.e., $(\mathcal{T}^{av,k})'=\mathcal{T}^{k+1}$, which can be computed as follows:
\begin{equation}
    \label{eq:mpcat1_text}
    \begin{split}
    &\mathcal{T}^{k+1}=\mathrm{NM}(\mathcal{T}^{av,k}+\mathrm{FF}(\mathcal{T}^{av,k})),\\
    &\mathbf{s.t.}\ \mathrm{FF}(\mathcal{T}^{av,k})=\\
    &\quad\ \ \mathrm{DP}(\mathrm{W}_{1}^k(\mathrm{DP}(\rho(\mathrm{W}_{0}^k\mathcal{T}^{av,k}+\mathrm{b}_{0}^k)))+\mathrm{b}_{1}^k),
    \end{split}
\end{equation}
where $\mathrm{NM}$, $\mathrm{FF}$, and $\mathrm{DP}$ is the normalization, feedforward and dropout layers, respectively; $\rho$ denotes non-linear activation function, e.g., $\mathrm{Relu}$; $\mathrm{W}_{0}^k$, $\mathrm{W}_{1}^k$, $\mathrm{b}_{0}^k$, and $\mathrm{b}_{1}^k$ are the learnable parameters.

Likewise, we can follow the processing of $(\mathcal{T}^{av,k})'$ to obtain the $k$-th layer acoustic feature matrix $(\mathcal{A}^{tv,k})'$, a.k.a., $\mathcal{A}^{k+1}$, and visual feature matrix $(\mathcal{V}^{ta,k})'$, a.k.a., $\mathcal{V}^{k+1}$. $\mathcal{A}^{k+1}$ and $\mathcal{V}^{k+1}$ are encoded by two MPCATs. $\mathcal{A}^{k+1}$ and $\mathcal{V}^{k+1}$ are formulated as follows:
\begin{equation}
    \label{eq:mpcat_acoustics}
    \begin{split}
    &\mathcal{A}^{tv,k}=\mathrm{NM}(\mathcal{A}^k+\mathrm{MPA}(\mathcal{T}^k,\mathcal{V}^k,\mathcal{A}^k)),\\
    &\mathcal{A}^{k+1}=\mathrm{NM}(\mathcal{A}^{tv,k}+\mathrm{FF}(\mathcal{A}^{tv,k})),
    \end{split}
\end{equation}
\begin{equation}
    \begin{split}
        &\mathcal{V}^{ta,k}=\mathrm{NM}(\mathcal{V}^k+\mathrm{MPA}(\mathcal{T}^k,\mathcal{A}^k,\mathcal{V}^k)),\\
        &\mathcal{V}^{k+1}=\mathrm{NM}(\mathcal{V}^{ta,k}+\mathrm{FF}(\mathcal{V}^{ta,k})),
    \end{split}
\end{equation}
where $\mathcal{A}^{tv,k}$ is the $k$-th layer acoustic feature matrix that are interacted by textual and visual modalities; similarly, $\mathcal{V}^{ta,k}$ is the $k$-th layer visual feature matrix that are interacted by textual and acoustic modalities.

After $\mathcal{T}$, $\mathcal{A}$, $\mathcal{V}$ are encoded by the $k$-th layer MPCATs, we obtain three new feature matrices: $\mathcal{T}^{k+1}$, $\mathcal{A}^{k+1}$, $\mathcal{V}^{k+1}$, which corresponding to textual, acoustic and visual modalities, respectively. Here, $\mathcal{T}^{0}=\mathrm{X}^{t,L}$, $\mathcal{A}^{0}=\mathrm{X}^{a,L}$, $\mathcal{V}^{0}=\mathrm{X}^{v,L}$, and $L$ denotes the number of layers of MDGATs. These new feature matrices are results of the $k$-th layer cross-modal feature interactions. It is worth noting that $\mathcal{T}^{k+1}$, $\mathcal{A}^{k+1}$, and $\mathcal{V}^{k+1}$ are not only the outputs of the $k$-th layer MPCATs but also the inputs of the $(k+1)$-th layer MPCATs.

\subsection{Multimodal Emotion Classification}
After the $K$ layers MPCATs encoding, we obtain the last layer feature matrices $\mathcal{T}^{(K)}$, $\mathcal{A}^{(K)}$, $\mathcal{V}^{(K)}$. We concatenate $\mathcal{T}^{(K)}$, $\mathcal{A}^{(K)}$, $\mathcal{V}^{(K)}$ to obtain final feature matrix $Z$ containing the information about three modalities. The concatenation operation can be formalized as follows:
\begin{equation}
    \label{eq:fusion}
    Z = \mathrm{W}_{u}[\mathcal{T}^{(K)} \ \Vert \ \mathcal{A}^{(K)} \ \Vert \ \mathcal{V}^{(K)}],
\end{equation}
where $\Vert$ is the concatenation operation; $\mathrm{W}_{u}$ is the learnable parameter. Then $Z$ is used as the input of multimodal classification module to predict emotional states of utterances. The emotion prediction can be formalized as follows:
\begin{equation}
    \label{eq:predict}
    \begin{split}
        &h_i = \mathrm{ReLU}(\mathrm{W}_{c}z_i+\mathrm{b}_{c}),\\
        &p_i = \mathrm{Softmax}(\mathrm{W}'_{c}h_i+\mathrm{b}'_{c}),\\
        &\hat{y}_i = \mathop{\mathrm{argmax}}_\mathrm{q}(p_i[\mathrm{q}]),
    \end{split}
\end{equation}
where $z_i \in Z$ is the feature vector of the $i$-th utterance $u_i$; $p_i$ denotes the probability distribution of predicted emotion label of $u_i$, and $\hat{y}_i$ denotes the predicted emotion; $\mathrm{ReLU}$ and $\mathrm{Softmax}$ denote the activation function and softmax function, respectively; $\mathrm{W}_{c}$, $\mathrm{W}'_{c}$, $\mathrm{b}_{c}$, and $\mathrm{b}'_{c}$ are the trainable parameters.

\subsection{Training Objective}
Finally, we adopt cross-entropy with L2-regularization as objective function to train the proposed GA2MIF, which can be represented as follows:
\begin{equation}
    \label{eq:loss}
    \mathcal{L} = - \frac {1}{\sum_{t=0}^{N-1} n(t)} \sum_{i=0}^{N-1}\sum_{j=0}^{n(i)-1} y_{ij} \log p_{ij} + \eta  \lvert \mathrm{W}_{l} \rvert,
\end{equation}
where $n(i)$ is the number of utterances of the $i$-th conversation, and $N$ is the number of all conversations in training set; $p_{ij}$ denotes the probability distribution of predicted emotion label of the $j$-th utterance in the $i$-th conversation, and $y_{ij}$ denotes the ground truth label of the $j$-th utterance in the $i$-th conversation; $\eta$ is the L2-regularizer weight, and $\mathrm{W}_{l}$ is the set of all learnable parameters. 

\section{Experimental Settings}\label{experimental_settings}
\subsection{Dataset Descriptions}
We conduct experiments for the multimodal ERC task on the two extensively adopted datasets: Interactive Emotional Dyadic Motion Capture (IEMOCAP) \cite{busso2008iemocap} and Multimodal Emotion Lines Dataset (MELD) \cite{poria2018meld}. For the purpose of comparison, we follow the methods of Hu et al. \cite{hu2021mmgcn}: the textual features are extracted by adopting TextCNN \cite{kim-2014-convolutional}; the acoustic features are extracted through using OpenSmile toolkit \cite{schuller2011recognising}; and the visual features are extracted by employing DenseNet \cite{huang2017densely}. The statistics of two datasets are shown in TABLE~\ref{tab:statistics}.
\begin{table*}[htbp]
    \centering
    \renewcommand{\arraystretch}{1.0}
    \setlength{\tabcolsep}{3pt}
    \caption{The Statistics of Two Datasets We Used. Here, \#Conversations, \#Utterances, and \#Classes denote the number of conversations, utterances, and classes on the datasets, respectively; \#Utterances per conversation, \#Speakers per conversation denote the number of utterances and speakers in each conversation, $Avg.$ indicates the corresponding average number}
    \begin{tabular}{c|ccc|ccc|c|ccc|c}
    \hline
    \multirow{2}[0]{*}{Dataset} & \multicolumn{3}{c|}{\#Conversations} & \multicolumn{3}{c|}{\#Utterances} & \multirow{2}[0]{*}{\#Classes} & \multicolumn{3}{c|}{\#Utterances per conversation} & \multirow{2}[0]{*}{\#Speakers per conversation} \\  
            & \textit{train} & \textit{valid} & \textit{test} & \textit{train} & \textit{valid} & \textit{test} &  & \textit{train} & \textit{valid} & \textit{test}&  \\
    \hline
    IEMOCAP & \multicolumn{2}{c}{120} & 31    & \multicolumn{2}{c}{5810} & 1623  & 6  & \multicolumn{2}{c}{48.39 ($Avg.$)}   & 52.32 ($Avg.$)&  2   \\
    MELD  & 1039  & 114   & 280   & 9989  & 1109  & 2610  & 7  & 9.6 ($Avg.$)  & 9.7 ($Avg.$)   & 9.3 ($Avg.$)&  3 or more  \\
    \hline
    \end{tabular}%
    \label{tab:statistics}%
\end{table*}

\textbf{IEMOCAP} dataset is a dyadic multimodal ERC dataset and contains audio-video-text data of impromptu performances or scripted scenes of about ten participants. Each video includes a single conversation, and each conversation consists of multiple utterances. There are in total 151 conversations and 7433 utterances on the IEMOCAP dataset, and each utterance corresponds to an emotional label. In IEMOCAP dataset, there are six emotion labels in total, including \textit{Happy}, \textit{Sad}, \textit{Neutral}, \textit{Angry} \textit{Excited}, and \textit{Frustrated}. \textbf{MELD} dataset is a multi-party and multimodal conversational emotion recognition dataset collected from the TV show \textit{Friends}. There are seven emotion categories containing \textit{Neutral}, \textit{Surprise}, \textit{Fear}, \textit{Sadness}, \textit{Joy}, \textit{Disgust}, and \textit{Angry}. MELD includes the information of textual, acoustic and visual modalities with more than 1400 dialogues and 13000 utterances. Unlike the two-person conversation scenario on the IEMOCAP dataset, the MELD dataset has three or more participants in each conversation.

\subsection{Comparison Models}
In order to perform comprehensive evaluations of the proposed GA2MIF, we compare it to baseline models. These baselines include unimodal ERC model and multimodal ERC model, which are described as follows.

\textbf{MFN} \cite{zadeh2018memory} performs multi-views information fusion and unifies the features of different modalities, but it does not consider either context-aware dependencies or speaker-aware dependencies. \textbf{BC-LSTM} \cite{poria2017context} employs textual, visual and acoustic modalities for multimodal ERC task, and adopts an utterance level LSTM to capture multimodal information. \textbf{CMN} \cite{hazarika2018conversational} leverages multimodal information by directly concatenating the features from three modalities, and uses GRU for contextual modeling. \textbf{ICON} \cite{hazarika2018icon} models the contextual knowledge of self- and inter-speaker influences through a GRU-based multi-hop memory network. \textbf{DialogueRNN} \cite{majumder2019dialoguernn} recognizes current emotion by tracking the contextual information of utterance and taking the characteristic of speaker into account. \textbf{DialogueGCN} \cite{ghosal2019dialoguegcn} is a graph-based model that regards the current conversation as a graph, where nodes represent utterances in that conversation. DialogueGCN can effectively capture long-range contextual information. In order to implement multimodal setting, we directly concatenate features of three modalities. \textbf{DialogueCRN} \cite{hu2021dialoguecrn} is designed with multi-turn reasoning modules to extract and integrate affective cues and can sufficiently understand the contextual information from a cognitive perspective. We implement multimodal setting for DialogueCRN through directly concatenating features of three modalities. \textbf{MMGCN} \cite{hu2021mmgcn} is a multimodal ERC model that simultaneously captures both intra-modal contextual information and inter-modal interactive information.

\subsection{Implementation Details}
All of our experiments are run on NVIDIA GeForce RTX 3080 Ti. All models are implemented via the PyTorch toolkit, as well as the maximum epoch is set to 100. The optimizer used for all models is AdamW, the L2 regularization factor is 0.00001, and the dropout rate is 0.1. $\mathrm{Norm}$ is replaced by the layer normalization in GA2MIF. The settings of partial hyperparameters are shown in TABLE~\ref{tab:implementation}. For IEMOCAP dataset, the number of layers of MDGATs is 3, and that of MPCATs is 4; the trade-off parameter of speaker embedding $\lambda$ is 1.6; the learning rate is 0.00001; the batch size is 8. For MELD dataset, the number of layers of MDGATs is 2, and that of MPCATs is 2; the trade-off parameter of speaker embedding $\lambda$ is 0.6; the learning rate is 0.00001; the batch size is 32. For a more rigorous comparison with the optimal baseline MMGCN, our ratios of the training, validation, and test sets are aligned with those of MMGCN. Referring to previous works \cite{ghosal2019dialoguegcn,hu2021mmgcn}, we evaluate the performance of GA2MIF with weighted-average F1 score and average accuracy.
\begin{table}[htbp]
    \centering
    \renewcommand{\arraystretch}{1.0}
    \setlength{\tabcolsep}{1.5pt}
    \caption{Description of the Partial Hyperparameters}
    \begin{tabular}{c|cc|c|c|c}
    \hline
    \multirow{2}[0]{*}{Dataset} & \multicolumn{2}{c|}{Number of layers} & \multirow{2}[0]{*}{\makecell{Trade-off \\parameter $\lambda$}} & \multirow{2}[0]{*}{Learning rate} & \multirow{2}[0]{*}{Batch size}\\
            & MDGATs & MPCATs &  & \\
    \hline
    IEMOCAP &  3  & 4 &  1.6  & 1e-05 & 8\\
    MELD  & 2  & 2 & 0.6 & 1e-05 & 32\\
    \hline
    \end{tabular}%
    \label{tab:implementation}%
\end{table}

\section{Comparison and Analysis}\label{comparison_analysis}
To illustrate the validity of our model, we conduct extensive experiments in this section. First of all, we intuitively compare the experimental results of our model with those of baseline models. Then, we discuss the effect of different settings on proposed GA2MIF; All experimental results are reported with the help of tables and figures. Finally, we provide two case studies in the last part of this section.

\subsection{Comparison With Baseline Models}
\begin{table*}[htbp]
    \centering
    \renewcommand{\arraystretch}{1.0}
    \setlength{\tabcolsep}{3.5pt}
    \caption{Overall Results of All Models on Both IEMOCAP and MELD Datasets. Evaluation metrics contain $\mathrm{Acc}$, $\mathrm{F1}$, and $\mathrm{wa}$-$\mathrm{F1}$, which denote accuracy score (\%), F1 score (\%), and weighted-average F1 score (\%). Best performance in bold}
    \begin{tabular}{c|cccccc|cc||ccccc|cc}
    \hline
    \multicolumn{1}{c|}{\multirow{3}{*}{Model}} & \multicolumn{8}{c||}{IEMOCAP} & \multicolumn{7}{c}{MELD} \\\cline{2-16}          
    & \textit{Happy} & \textit{Sad}   & \textit{Neutral} & \textit{Angry} & \textit{Excited} & \textit{Frustrated} & \multirow{2}{*}{Acc} & \multirow{2}{*}{wa-F1} & \textit{Neutral} & \textit{Surprise} & \textit{Sadness} & \textit{Joy} & \textit{Anger} & \multirow{2}{*}{Acc} & \multirow{2}{*}{wa-F1} \\ \cline{2-7} \cline{10-14}
    & F1 & F1   & F1 & F1 & F1 & F1 &  &  & F1 & F1 & F1 & F1 & F1 &  &  \\
    \hline
      MFN & 47.19 & 72.49 & 55.38 & 63.04 & 64.52 & 61.91 & 60.14 & 60.32  & 76.28 & 48.29 & 23.27 & 52.23  & 41.32 & 59.93 & 57.29 \\
      BC-LSTM & 32.63 & 70.34 & 51.14 & 63.44 & 67.91 & 61.06 & 59.58 & 59.10  & 75.66 & 48.47 & 22.06 & 52.10  & 44.39 & 59.62 & 56.80 \\
      CMN   & 30.38 & 62.41 & 52.39 & 59.83 & 60.25 & 60.69 & 56.56 & 56.13 &  - & - & - & - & - & - & - \\
      ICON  & 29.91 & 64.57 & 57.38 & 63.04 & 63.42 & 60.81 & 59.09 & 58.54 &  - & - & - & - & - & - & - \\
      DialogueRNN & 33.18 & 78.80  & 59.21 & 65.28 & 71.86 & 58.91 & 63.40  & 62.75 & 76.79 & 47.69 & 20.41 & 50.92 & 45.52 & 60.31 & 57.66 \\
      DialogueGCN & 47.10  & 80.88 & 58.71 & 66.08 & 70.97 & 61.21 & 65.54 & 65.04 & 75.97 & 46.05 & 19.60  & 51.20  & 40.83 & 58.62 & 56.36 \\
      DialogueCRN & \textbf{51.59} & 74.54 & 62.38 & 67.25 & 73.96 & 59.97 & 65.31 & 65.34 & 76.13 & 46.55 & 11.43 & 49.47 & 44.92 & 59.66 & 56.76 \\
      MMGCN & 45.45 & 77.53 & 61.99 & 66.67 & 72.04 & 64.12 & 65.56 & 65.71 & 75.16 & 48.45 & 25.71 & \textbf{54.41} & 45.45 & 59.31 & 57.82 \\
      \hline
      GA2MIF  & 46.15 & \textbf{84.50} & \textbf{68.38} & \textbf{70.29}  & \textbf{75.99} & \textbf{66.49} & \textbf{69.75} & \textbf{70.00} & \textbf{76.92} & \textbf{49.08} & \textbf{27.18} & 51.87 & \textbf{48.52} & \textbf{61.65} & \textbf{58.94} \\
      \hline
    \end{tabular}%
    \label{tab:overall}%
\end{table*}%
\begin{figure*}[htbp]
    \centering
    \subfloat[Confusion matrix of MMGCN]{\includegraphics[width=3.5in]{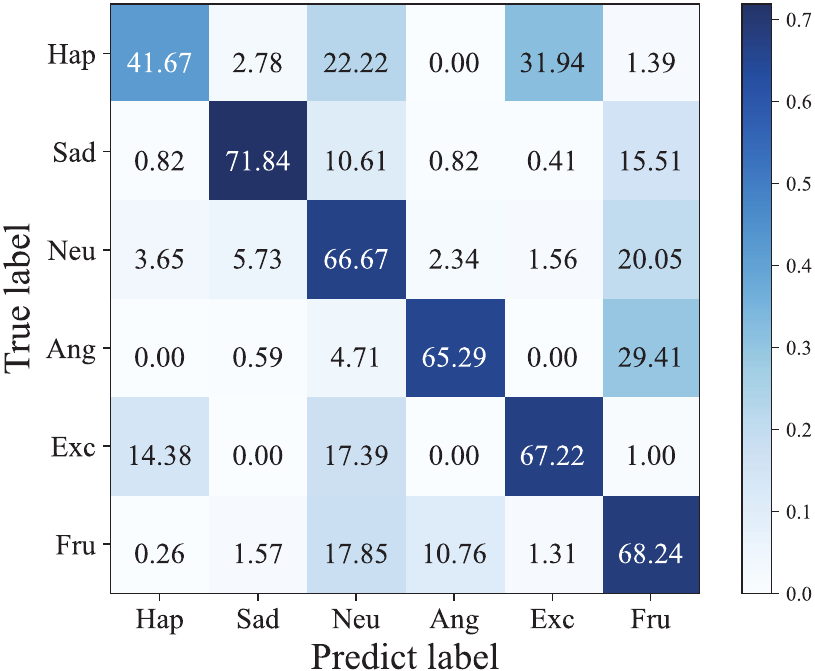}%
    \label{fig:cm_mmgcn}}
    \hfil
    \subfloat[Confusion matrix of GA2MIF]{\includegraphics[width=3.5in]{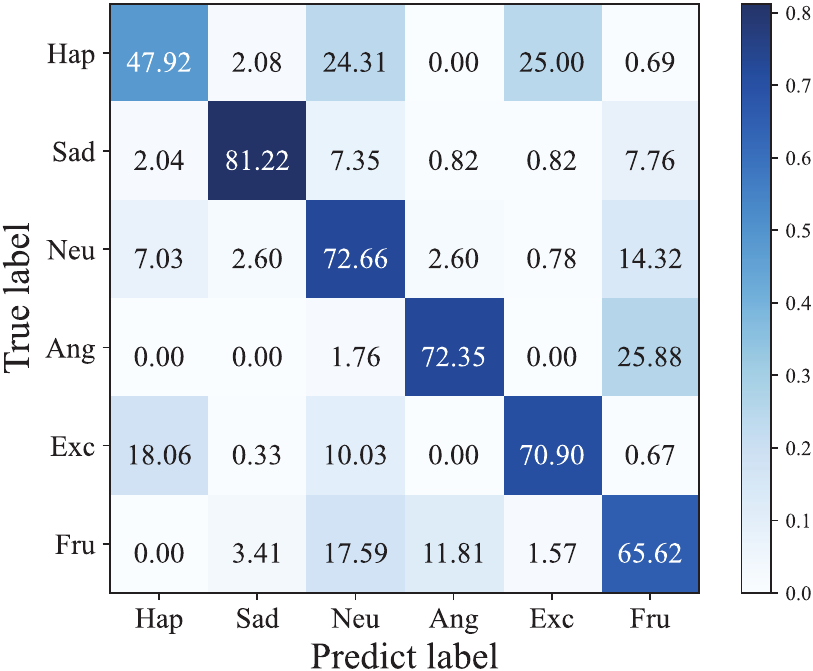}%
    \label{fig:cm_ga2mif}}
    \caption{Comparison of confusion matrices between MMGCN and GA2MIF. Hap, Sad, Neu, Ang, Exc, Fru denote \textit{Happy}, \textit{Sad}, \textit{Neutral}, \textit{Angry}, \textit{Excited}, and \textit{Frustrated}, respectively.}
    \label{fig:cm}
\end{figure*}
TABLE~\ref{tab:overall} reports the experimental results of GA2MIF with other baseline models. It can be visualized from TABLE~\ref{tab:overall} that our GA2MIF achieves optimal accuracy scores and weighted-average F1 scores on both IEMOCAP and MELD datasets. For the IEMOCAP dataset, the accuracy and F1 score of GA2MIF are 69.75\% and 70.00\%, which are 4.19\% and 4.29\% higher than those of the strongest baseline model, i.e., MMGCN. For the MELD dataset, likewise, the accuracy and F1 score of our GA2MIF respectively increase by 2.34\% and 1.12\% in comparison with those of MMGCN. We can conclude from above results that the improvements of GA2MIF on the MELD dataset are not as significant as that on the IEMOCAP dataset. After carefully comparing the differences between IEMOCAP and MELD, we notice that two neighboring utterances on the MELD dataset may not be consecutive sentences in real conversation scenarios. Therefore, GA2MIF is identical to most models on the MELD dataset, and it is challenging to take advantage of model itself in the absence of a powerful feature extractor.

We report F1 score corresponding to each emotion label in detail in TABLE~\ref{tab:overall}. Our GA2MIF obtain the highest F1 scores on the IEMOCAP dataset for most emotion labels except for \textit{Happy}. We can observe that the results of GA2MIF show remarkably significant improvement relative to those of baseline models for \textit{Sad} and \textit{Neutral}. Fig.~\ref{fig:cm} depicts confusion matrix of GA2MIF and MMGCN on the IEMOCAP dataset. Fig.~\ref{fig:cm_ga2mif} indicates that our GA2MIF recognizes \textit{Sad} better than other emotion labels, with an accuracy score of 81.22\%. By comparing Fig.~\ref{fig:cm_mmgcn} and Fig.~\ref{fig:cm_ga2mif}, we can conclude that the accuracy scores of GA2MIF are higher than those of MMGCN except for \textit{Frustrated}.

\subsection{Impact of Different Trade-Off Parameters}
\begin{figure*}[htbp]
  \centering
  \subfloat[Accuracy and F1 scores on the IEMOCAP dataset]{\includegraphics[width=3.5in]{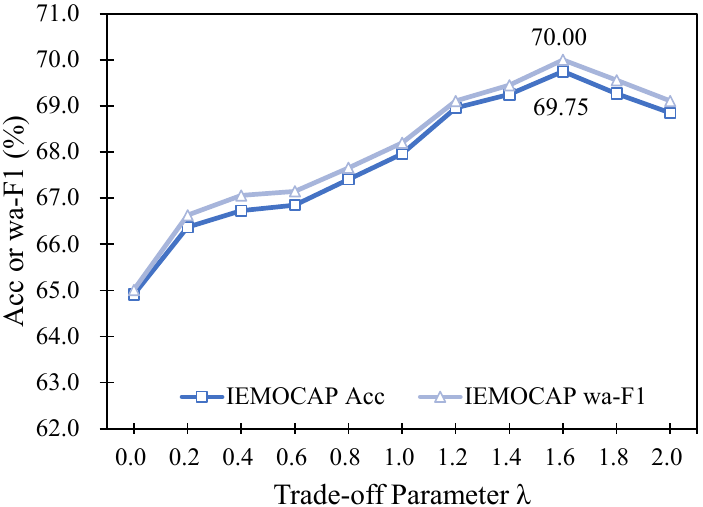}%
  \label{fig:speaker0}}
  \hfil
  \subfloat[Accuracy and F1 scores on the MELD dataset]{\includegraphics[width=3.5in]{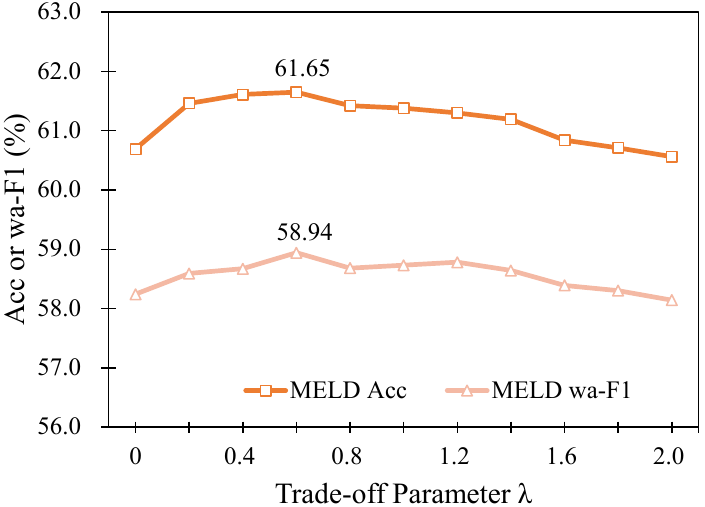}%
  \label{fig:speaker1}}
  \caption{The performance of our model with different trade-off parameters of speaker embedding. $\mathrm{Acc}$ and $\mathrm{wa}$-$\mathrm{F1}$ denote the accuracy and weighted-average F1 score, respectively; $\lambda$ indicates the trade-off parameter of speaker embedding.}
  \label{fig:speaker}
\end{figure*}
Fig.~\ref{fig:speaker} shows the influence of different trade-off parameters of speaker embedding on the results. Fig.~\ref{fig:speaker0} shows results on the IEMOCAP dataset, while the results on the MELD dataset are shown in Fig.~\ref{fig:speaker1}. We find from Fig.~\ref{fig:speaker0} that the performance of GA2MIF gradually increases as the trade-off parameter increases. This phenomenon suggests that the characteristic of speaker plays an essential role in emotion recognition task. It is noteworthy that the performance of GA2MIF starts to decrease when increasing to a certain threshold (i.e., $\lambda=1.6$). As shown in Fig.~\ref{fig:speaker1}, the performance variation of GA2MIF on the MELD dataset is generally consistent with trend on the IEMOCAP dataset. In other words, the values of accuracy score and weighted-average F1 score increase with the increasing of trade-off parameter $\lambda$ within a certain margin (i.e., $\lambda<0.6$).

\subsection{Impact of Different Window Sizes in Directed Graph}
\begin{figure*}[htbp]
    \centering
    \subfloat[Accuracy and F1 scores on the IEMOCAP dataset]{\includegraphics[width=3.5in]{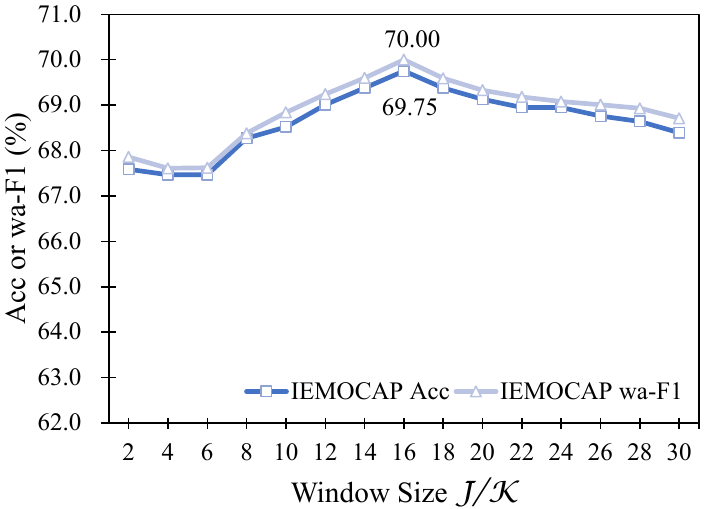}%
    \label{fig:window0}}
    \hfil
    \subfloat[Accuracy and F1 scores on the MELD dataset]{\includegraphics[width=3.5in]{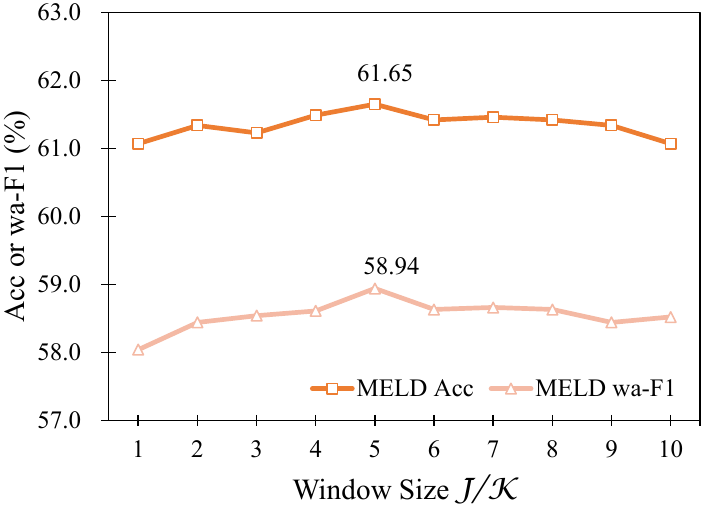}%
    \label{fig:window1}}
    \caption{The accuracy and weighted-average F1 scores of our GA2MIF with different window sizes in directed graph.}
    \label{fig:window}
\end{figure*}
In this part, we discuss the effect of different window sizes on the performance of GA2MIF. Fig.~\ref{fig:window} shows the results of GA2MIF corresponding to different window sizes on the IEMOCAP dataset. As shown in Fig.~\ref{fig:window0}, we set various window sizes (i.e., $(2,2)$, $(4,4)$, $\cdots$, $(30,30)$) for GA2MIF on the IEMOCAP dataset, each of which is a combination of $\mathcal{J}$ and $\mathcal{K}$. As we expected, the accuracy score and weighted-average F1 score increase with the increasing of the window size within a certain margin. When the window size increases to a certain threshold, i.e., $(16,16)$, the performance of GA2MIF gradually starts to decrease. In the same way, we set different window sizes (i.e., $(1,1)$, $(2,2)$, $\cdots$, $(10,10)$) on the MELD dataset, and obtain conclusions that are similar to those on the IEMOCAP dataset. On the MELD dataset, however, the effect of window sizes on the performance of GA2MIF is relatively slight. The results on the MELD dataset are shown in Fig.~\ref{fig:window1}.

\subsection{Impact of Different Updating Functions}
\begin{table}[htbp]
    \centering
    \renewcommand{\arraystretch}{1.2}
    \setlength{\tabcolsep}{7pt}
    \caption{The Performance of GA2MIF Under Various Updating Functions in MDGATs. The updating functions include $Sum$, $Concat$, and $Sum$-$Product$, respectively}
      \begin{tabular}{c|cc||cc}
      \hline
      \multicolumn{1}{c|}{\multirow{2}[1]{*}{Updating Function}} & \multicolumn{2}{c||}{IEMOCAP}  & \multicolumn{2}{c}{MELD} \\
      \cline{2-5}
            & Acc & wa-F1 & Acc & wa-F1 \\
      \hline
      Sum     & 68.94  & 68.98  & 61.19  & 58.70  \\
      Concat     & 68.91  & 68.95  & 61.19  & 58.68  \\
      \hline
      Sum-Product & \textbf{69.75}  & \textbf{70.00}  & \textbf{61.65} & \textbf{58.94} \\
      \hline
      \end{tabular}%
    \label{tab:updating}%
\end{table}%
We design a novel updating function, \textit{Sum-Product}, for MDGATs in Section \ref{mdgats}. TABLE~\ref{tab:updating} shows the results of proposed GA2MIF adopting \textit{Sum}, \textit{Concat} and \textit{Sum-Product} functions. From the experimental results, we can observe that \textit{Sum-Product} we designed has a slight improvement over the other two updating functions. In our future work, we hope to apply \textit{Sum-Product} to feature interaction-sensitive tasks, such as knowledge graph-based emotion recognition, sentiment analysis-based recommendation systems.

\subsection{Impact of Different Number of Network Layers}
\begin{figure*}[htbp]
    \centering
    \subfloat[Variation of accuracy with different number of layers]{\includegraphics[width=3.5in]{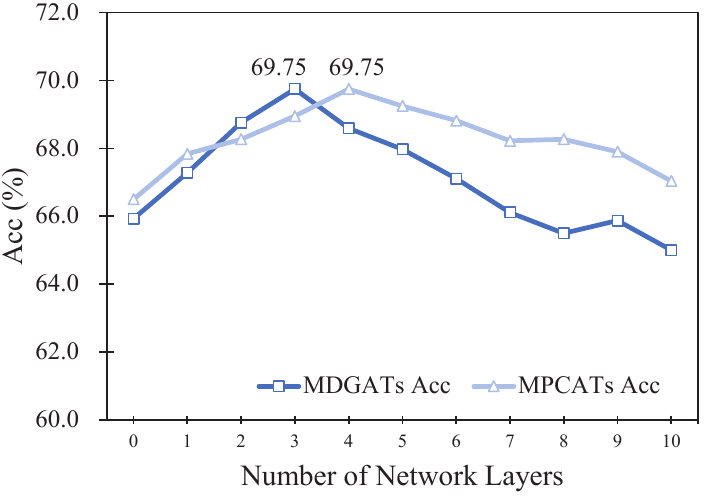}%
    \label{fig:layer0}}
    \hfil
    \subfloat[Variation of F1 score with different number of layers]{\includegraphics[width=3.5in]{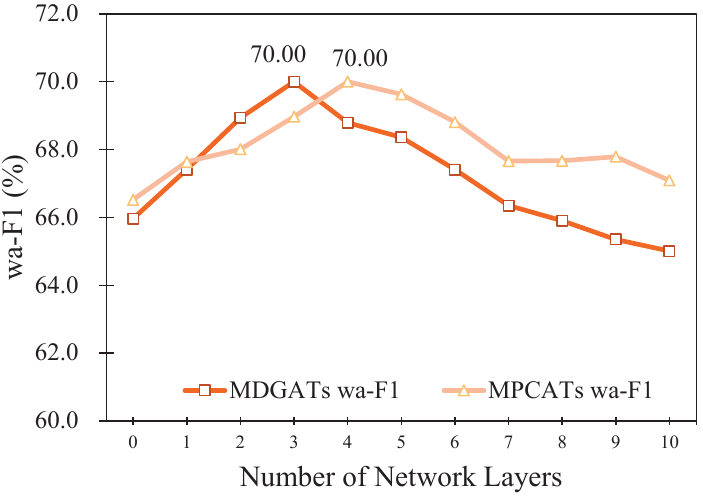}%
    \label{fig:layer1}}
    \caption{Effect of different number of network layers on the performance of GA2MIF. (a) The dark (or light) blue line denotes the effect of different layers in MDGATs (or MPCATs) on accuracy scores. (b) The dark (or light) orange line denotes the effect of different layers in MDGATs (or MPCATs) on weighted-average F1 scores.}
    \label{fig:layer}
\end{figure*}
Our GA2MIF mainly consists of two sub-networks, namely MDGATs and MPCATs. We depict the effect of different number of layers in MDGATs or MPCATs on the performance of GA2MIF in Fig.~\ref{fig:layer}. The discussions and analyses in this subsection are based on the experimental results of the IEMOCAP dataset. Fig.~\ref{fig:layer0} illustrates the effect of different number of network layers on the accuracy scores. Notably, before discussing how the performance is affected by the number of layers in current sub-network, we fix the number of layers in the other sub-network. As shown by the dark blue line in Fig.~\ref{fig:layer0}, we first set the number of layers in MPCATs to 4 and then plot the variation curve of accuracy score under different number of layers in MDGATs; conversely, the number of layers in MDGATs is first limited to 3, and then the change of accuracy score under different number of layers in MPCATs is recorded by the light blue line in Fig.~\ref{fig:layer0}. It can be found that with the increase of layers in MDGATs, the accuracy score of GA2MIF rises first and then falls. The effect of different number of layers in MPCATs on accuracy score also follows this pattern, i.e., the accuracy score increases first and then decreases as the number of layers in MPCATs increases. The difference is that the number of layers in MDGATs is more sensitive to the performance of GA2MIF than that in MPCATs. Similarly, Fig.~\ref{fig:layer1} shows the effect of different number of network layers on the weighted-average F1 scores. We can draw the analogous conclusion from Fig.~\ref{fig:layer1} as Fig.~\ref{fig:layer0}. As shown in Fig.~\ref{fig:layer}, it is noteworthy that the performance of GA2MIF shows a significant degradation when we do not use either MDGATs (i.e., MDGATs with 0 layers) or MPCATs (i.e., MPCATs with 0 layers). Therefore, MDGATs and MPCATs can contribute to the performance of our model.

\subsection{Performance under Different Modality Settings}
\begin{table}[htbp]
    \centering
    \renewcommand{\arraystretch}{1.2}
    \setlength{\tabcolsep}{7pt}
    \caption{The Performance of GA2MIF Under Different Modality Settings. $a$, $v$, $t$ indicate acoustic, visual, and textual modalities, respectively}
      \begin{tabular}{c|cc||cc}
      \hline
      \multicolumn{1}{c|}{\multirow{2}[1]{*}{Modality Setting}} & \multicolumn{2}{c||}{IEMOCAP}  & \multicolumn{2}{c}{MELD} \\
      \cline{2-5}
         & Acc & wa-F1 & Acc & wa-F1 \\
      \hline
    $a$ + $v$ & 59.70  & 59.85  & 48.12  & 43.34  \\
    $a$ + $t$ & 67.47  & 67.49  & 60.04  & 57.11  \\
    $v$ + $t$ & 64.26  & 64.39  & 60.00  & 57.22  \\
      \hline
    $a$ + $v$ + $t$ & \textbf{69.75}  & \textbf{70.00}  & \textbf{61.65} & \textbf{58.94} \\
      \hline
      \end{tabular}%
    \label{tab:modality-setting}%
\end{table}%
In this subsection, we conduct experiments with two- and three-modality settings on two public datasets. Note that the proposed GA2MIF requires at least two modalities. These modality settings include the acoustic-visual setting, acoustic-textual setting, visual-textual setting, and acoustic-visual-textual setting. As shown in TABLE~\ref{tab:modality-setting}, we report accuracy and weighted-average F1 scores of all modality settings. It can be seen from TABLE~\ref{tab:modality-setting} that the experimental results of three-modality setting outperform those of all two-modality settings. Among the two-modality settings, our model achieves the best results under the acoustic-textual setting and the worst performance under the acoustic-visual setting. On the IEMOCAP dataset, the accuracy and F1 scores reach 67.47\% and 67.49\%, respectively, under the acoustic-textual setting, which are 7.77\% and 7.64\% higher than those under the visual-textual setting. Moreover, the results of GA2MIF under the acoustic-textual setting are higher than those of MMGCN under three-modality setting, which fully demonstrates the effectiveness of the proposed GA2MIF.

\subsection{Case Studies}
\begin{figure}[htbp]
    \centering
    \includegraphics[width=3.4in]{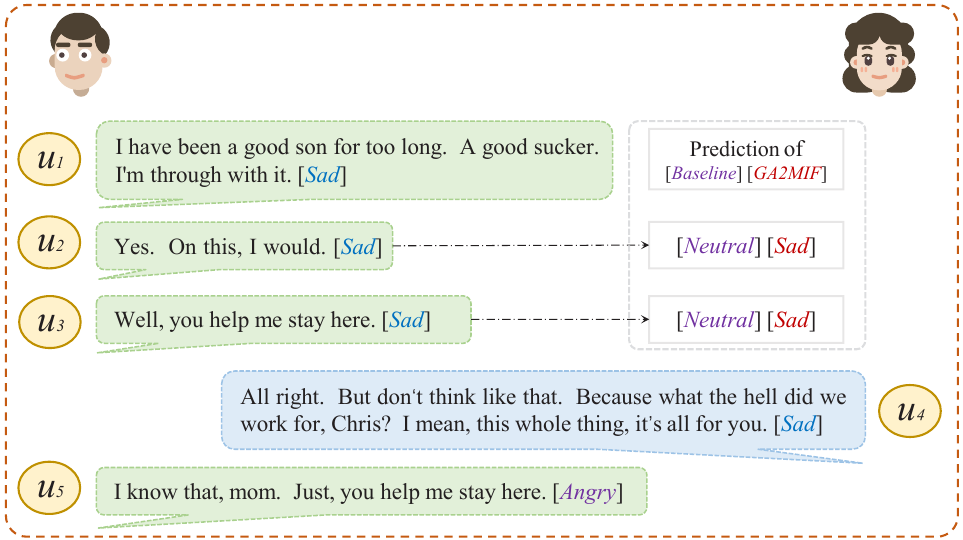}
    \caption{Example of emotion recognition in conversation on the IEMOCAP dataset. The proposed GA2MIF can correctly recognize non-\textit{Neutral} emotions.}
    \label{fig:case0}
\end{figure}

In the ERC task, several utterances with non-\textit{Neutral} emotion labels such as ``\textit{Yes.  On this, I would.} [\textit{Sad}]'', ``\textit{I do.} [\textit{Angry}]'', ``\textit{Actually, now that you mention it, no. I don't.} [\textit{Excited}]'', are difficult to be detected correctly by existing models. Most of existing text-modal ERC models tend to directly recognize these utterances as \textit{Neutral}. The above-mentioned scenario is shown in Fig.~\ref{fig:case0}. Intuitively, multimodal ERC models such as MMGCN can compensate the inadequacies of text-modal models by acoustic or visual modality. However, MMGCN treats utterances of all modalities as nodes of the same type, which clearly violates the assumption that the input of GNNs is a homogeneous graph and thus cannot effectively utilize multimodal information. Our GA2MIF inputs the information of each modality into MDGATs separately, and then employs MPCATs for inter-modal information interaction. Furthermore, our approach eliminates complex redundant connections by employing context window to connect edges instead of constructing fully connected graph, which allows for more efficient selection of useful contextual information. Experiments show that the proposed GA2MIF can detect emotional states more accurately relative to MMGCN.

\begin{figure}[htbp]
    \centering
    \includegraphics[width=3in]{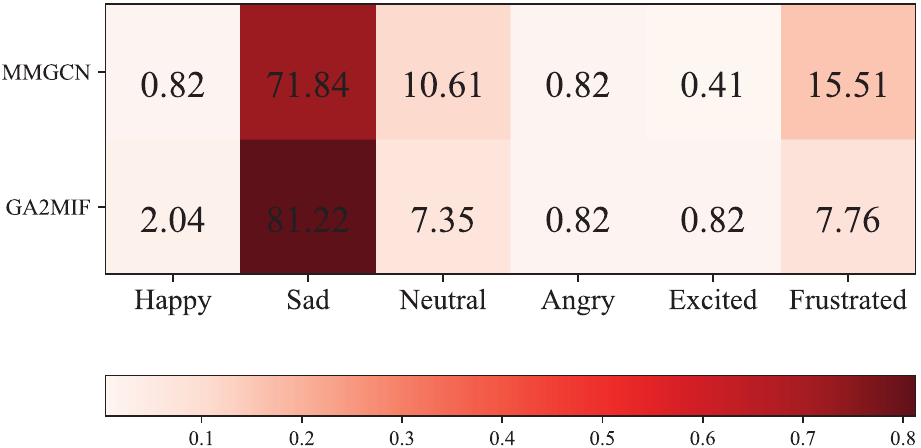}
    \caption{The probability of predicting ground-truth emotion \textit{Sad} as other emotions by MMGCN and GA2MIF.}
    \label{fig:case1}
\end{figure}
Like most baseline methods, the proposed GA2MIF has challenges in distinguishing similar emotions, e.g., \textit{Happy} vs \textit{Excited}, \textit{Sad} vs \textit{Frustrated}. Fortunately, our GA2MIF alleviates the problem of similar emotions to a certain extent. The probability of predicting ground-truth emotion \textit{Sad} as other emotions by MMGCN and GA2MIF is shown in Fig.~\ref{fig:case1}. We can derive from Fig.~\ref{fig:case1} that our GA2MIF identifies \textit{Sad} as \textit{Frustrated} on the IEMOCAP dataset with a probability of 7.76\%, while MMGCN with a probability of 15.51\%. Thus, the proposed GA2MIF achieves a more outstanding result with a 7.75\% reduction relative to MMGCN.

\section{Summary and Prospect}\label{summary_prospect}
We propose a novel multimodal conversational emotion recognition model, i.e., Graph and Attention based Two-stage Multi-source Information Fusion (GA2MIF), in this paper. The proposed GA2MIF is mainly inspired by graph attention network and multi-head attention mechanism. Multi-head Directed Graph ATtention networks (MDGATs) and Multi-head Pairwise Cross-modal ATtention networks (MPCATs) are designed to model intra-modal contexts and inter-modal interactions, respectively. The collaboration of MDGATs and MPCATs can effectively address the challenge of MMGCN in handling the heterogeneous graph, as well as, the complex redundant connections are eliminated through the context window. In addition, we design a new update function, Sum-Product, for MDGATs (arguably, Graph Neural Networks). We have demonstrated the effectiveness of GA2MIF on two extensively used datasets and achieved an impressive weighted-average F1 score of 70.00\% on the IEMOCAP dataset, which overwhelmingly outperforms all baseline models. Not only that, we also discuss and analyze the effect of different settings on the performance of GA2MIF.

In future work, we will continue to promote multimodal learning, and focus on how to enhance the expressiveness of acoustic and visual modalities. Furthermore, we hope to apply our model to more multimodal fusion scenarios, as well as tackle the notorious problems of \textit{similar-emotion} and \textit{emotion-shifting} in conversational emotion recognition.

\bibliographystyle{IEEEtran}
\balance
\bibliography{ga2mif}

\end{document}